\documentclass[12pt]{article}  
\usepackage{graphicx}  
\usepackage{dcolumn}   
\usepackage{bm}        
\usepackage{amssymb}   
\usepackage{cite}
\def\beq{\begin{equation}}
\def\eeq{\end{equation}}

\def\bea{\begin{eqnarray}}
\def\eea{\end{eqnarray}}
\def\ba{\begin{array}}                  
\def\ea{\end{array}}

\begin{document}

\begin{center}
\noindent{\Large \bf Toroidal dipole moment of the lightest neutralino 
in the cMSSM}  
\\[15pt]
{\large Luis G. Cabral-Rosetti$^a$,  Myriam
  Mondrag{\'o}n$^b$,\\ Esteban Reyes-P\'erez$^b$}\\[15pt]
{\small 
 $^a$Unidad de Ciencias B{\'a}sicas, Departamento de Posgrado,
Centro Interdisciplinario de Investigaci\'on y Docencia en 
Educaci\'on T\'ecnica, CIIDET. Av. Universidad 282 Pte., Col. Centro,
A. Postal 752, C.P. 76000, Santiago de Quer\'etaro, Qro., M\'exico.
\\[3pt]
  \texttt{luis@nucleares.unam.mx}  \\[7pt]
$^b$  Instituto de F{\'\i}sica, Universidad Nacional Aut\'onoma de M{\'e}xico\\
  Apdo. Postal 20-364, M{\'e}xico 01000, D.F., M{\'e}xico\\[3pt]
  \texttt{myriam@fisica.unam.mx}\quad  
\texttt{esteban.reyesperez@gmail.com}
}
\end{center}

\begin{abstract}
  We study the toroidal dipole moment of the lightest neutralino in
  the constrained Minimal Supersymmetric Standard Model. The toroidal
  dipole moment is the only electromagnetic property of the
  neutralino. Since the neutralino is the LSP in many versions of the
  MSSM and therefore a candidate for dark matter, its characterization
  through its electromagnetic properties is important both for
  particle physics and for cosmology.  We perform a scan in the
  parameter space of the cMSSM and find that the toroidal dipole
  moment is different from zero, albeit very small, in all the
  parameter space, and reaches a value around $10^{-3}$ GeV$^{-2}$ in a
  particular region of the parameter space, well below experimental bounds.

\end{abstract}


\section{Introduction}

One of the best motivated extensions of the Standard Model (SM) is the Minimal
Supersymmetric Standard Model (MSSM), since, besides giving a solution
to the hierarchy problem, provides us with a good candidate for cold
dark matter (CDM), namely, the lightest neutralino.

There are currently several experiments under way, and more planned
for the future for direct and indirect detection of dark matter (DM)
(for recent reviews on dark matter direction see \cite{Szelc:2010zz,Schnee:2011ef}).
If detected, it will be necessary to discriminate between different
candidates. To this end, it will be important to characterize as much
as possible the different candidates.  The neutralino is at present
one of the best candidates for DM, and its electroweak properties can
give us some insight into its nature.  Because it is neutral, these
properties appear only radiatively.

Recently, there has been intense work on the electroweak properties of
dark matter since they might be relevant in the calculation of DM
decays and annihilations \cite{Sigurdson:2004zp,Ciafaloni:2010ti,Ciafaloni:2010qr,Ciafaloni:2011sa,Bell:2010ei,
Bell:2011eu,Bell:2011if,Dent:2008qy,Kachelriess:2009zy,Heo:2009xt,Heo:2009vt,Liebler:2010bi,Barger:2010gv},
 which have consequences in astrophysical processes
\cite{Berezinsky:2002hq,Kachelriess:2007aj} and therefore are important in indirect
astrophysical searches for DM, as in the calculation of the
annihilation cross section of the DM itself.

One of the least studied electromagnetic properties of a particle is
the Toroidal Dipole Moment (TDM), which is directly related to the
anapole moment.  The anapole moment corresponds to a $T$ invariant
interaction, which is $C$ and $P$ non-invariant \cite{Zeldovich:1957}.
It does not have a simple classical analogue, thus the toroidal dipole
moment was introduced as a more convenient description of $T$ even,
$C$ and $P$ odd, interactions
\cite{Afanasev:1991mb,Afanasev:1992qk,Dubovik:1996gx,Bukina:1998wn}.
The electromagnetic vertex of a particle can thus be expressed in a
multipole parametrization, including the toroidal moments (see for
instance \cite{Gongora:2006lk}), which provides a one to one correspondence
between the form factors and the multipole moments.

Pospelov and ter Veldhuis have obtained an upper limit for the anapole
moment of WIMPS \cite{Pospelov:2000bq}, using results from the DAMA
and CDMS experiments \cite{Bernabei:1996vj,Abusaidi:2000wg}.  In case
the neutralino is the main component of dark matter, its anapole
moment should comply with this limit.

In this paper we calculate the TDM of the neutralino within the
constrained Minimal Supersymmetric Standard Model.  We do a scan in
the five parameter space of the cMSSM, and compare the results with
the above mentioned experimental limit.  The article is organised as
follows: in section II we present a very brief summary of some aspects
of the constrained MSSM (cMSSM) relevant to our calculation.  In
section III we review the general form for the electromagnetic vertex
of a particle, and in particular for a Majorana particle, as the
neutralino.  We introduce the anapole moment and its relation to the
toroidal dipole moment.  In section IV we explain the methodology used
to calculate the TMD of the neutralino in the cMSSM and we evaluate
it for different values of the parameters.  Section V presents the
results obtained and our conclusions.

\section{The MSSM and the neutralino as candidate for dark matter}

The cold dark matter density is known to be \cite{Komatsu:2010fb}
\beq
\Omega_{DM}h^2 \sim 0.112 ~,
\eeq
where $h$ is the Hubble constant in units of $100$ km sec$^{-1}$ Mpc$^{-1}$.
The thermally averaged effective cross section times the relative
speed of the dark matter particle, needed to get this relic density is
\cite{Jungman:1995df,Olive:2003iq,Feng:2010gw}
\beq
<\sigma v> \propto g^4_{weak} /16\pi^2 m_x^2
\eeq
consistent with the assumption of a weakly interacting dark matter
particle (WIMP) with mass between $10 ~$GeV - $(few)$ TeV.

The minimal supersymmetric extension of the Standard Model (MSSM)
provides us with one of the best WIMP candidates for dark matter, namely the
lighest neutralino (for reviews on SUSY see for instance
\cite{Martin:1997ns, Aitchison:2005cf}).  The MSSM
requires two complex Higgs electroweak doublets to give mass to the up and
down type quarks and to avoid chiral anomalies. After  electroweak
symmetry breaking five physical Higgs states remain: two neutral CP
invariant $\left( h^{0}, H^{0}\right)$, two charged CP invariant $\left( H^{+}, H^{-}\right)$, and one neutral CP-odd $\left(
  A^{0}\right)$.

The MSSM has a new discrete symmetry, R parity, defined as 
 $R = \left( -1\right)^{3B+2S+L}$, where $B$ and $L$ are the baryonic
 and leptonic numbers respectively.  This symmetry assignes a charge
 +1 to the SM particles and -1 to the supersymmetric partners, thus
 making the lightest supersymmetric particle (LSP) stable. 

Supersymmetry has to be broken, or it would have already been
observed. To break supersymmetry explicitly, without the reappearance of quadratic divergencies, a set of super-renormalizable terms are added to the Lagrangian, the so-called soft breaking terms. The Lagrangian for the soft breaking terms is given by

\begin{equation} {\cal L}_{soft} = -\frac{1}{2}M_a\lambda^a\lambda^a
  -\frac{1}{6}A^{ijk}\phi_i\Phi_j\phi_k -
  \frac{1}{2}B^{ij}\phi_i\phi_j + 
  c.c. -(m^2)^i_j\phi^{j*}\phi_i ~,
\end{equation}

\noindent where $M_a$ are the gaugino masses, $A^{ijk}$ and $B^{ij}$ are
trilinear and bilinear couplings, respectively, and $(m^2)^i_j$ are
scalar squared-mass terms.  It is assumed that supersymmetry breaking
happens in a hidden sector, which communicates to the observable one
only through gravitational interactions, and that the gauge
interactions unifiy. This means that at the GUT scale the soft
breaking terms are ``universal'', i.e., 
the gauginos
$M_a$ have a common mass, as well as the scalars $(m^2)^i_j$ and the trilinear
couplings, $A^{ijk}$. Requiring electroweak symmetry breaking fixes
the value of $B^{ij}$ and the absolute value of the Higgsino mixing
parameter $|\mu|$. This is known as the constrained MSSM, or cMSSM, which
is described by five parameters: the unified gaugino mass
$m_{1/2}$, the universal scalar mass $m_0$, the value of the universal
trilinear coupling $A_0$, the sign of Higgsino mass parameter $\mu$, and the
ratio of the vacuum expectation values of the two Higgses, $\tan
\beta$.

After the electroweak symmetry breaking the neutral and charged states
in the MSSM can mix.  In the case of the neutral ones they give rise to
a set of four mass eigenstates, the neutralinos.  It is the lightest
one of these that is the LSP and a good candidate to dark matter in
many SUSY models.  The lightest neutralino, in the gauge eigenstate
basis, is thus a function of the neutral Higgsinos and the neutral
gauginos (Wino and Bino)
\begin{equation}
  \label{eq:neutralino}
  \psi_0 = (\tilde B,\tilde W^0, \tilde H^0_u, \tilde H^0_d) ~.
\end{equation}
The properties of the neutralinos will depend on the mixing, which in
turn depends on the soft breaking parameters.  Thus, the lightest
neutralino can range from almost pure Bino to almost pure Higgsino.

\section{Toroidal Dipole Moment}

For $1/2$-spin particles the most general expression for the
electromagnetic vertex function, which characterizes the interaction
between the particle and the electromagnetic field, is:

\begin{eqnarray}
\Gamma _\mu  (q) = f_Q (q^2 )\gamma _\mu   + f_\mu  (q^2 )i\sigma _{\mu \nu } q^\nu\gamma _5 \nonumber \\   - f_E (q^2 )\sigma _{\mu \nu } q^\nu   + f_A (q^2 )(q^2 \gamma _\mu   - \displaystyle{\not}q q_\mu )\gamma _5 ,
\end{eqnarray}

\noindent where $f_Q (q^2)$, $f_\mu (q^2)$, $f _E (q^2)$ and $f
_A(q^2)$ are the so called charge, magnetic dipole, electric dipole
and anapole form factors, respectively; where $q _{\mu} = p _{\mu} ' -
p_{\mu}$ is the transferred 4-momentum; and $\sigma _{\mu \nu} = (i/2)
\left[ {\gamma _\mu ,\gamma _\nu } \right]$ \cite{Dubovik:1996gx,
  Bukina:1998kw}. These form factors are physical observables when
$q^{2}\rightarrow0$, and their combinations define the well known
magnetic dipole $(\mu)$, electric dipole $(d)$ and anapole $(a)$ moments.

The electromagnetic properties of Majorana fermions (like the
neutralino) are described by a unique form factor, the anapole, $f_A
(q^2)$. This is a consequence of CPT-invariance and the C, P, T
properties of $\Gamma _\mu (q^{2})$ and the interaction Hamiltonian. Thus, the electromagnetic vertex function of a neutralino can be writen as
\begin{eqnarray}
  \Gamma _\mu  (q^2) =  f_A (q^2 )(q^2 \gamma _\mu   
- \displaystyle{\not}q q_\mu )\gamma _5 .
\end{eqnarray} 

The anapole moment was introduced by Zel'dovich to describe a
T-invariant interaction that does not conserve P and C parity
\cite{Zeldovich:1957}. The anapole moment does not have a simple
classical analogue, since $f_A(q^2)$ does not correspond to a
multipolar distribution. A more convenient quantity to describe this
interaction was proposed by V. M. Dubovik and A. A. Cheshkov
\cite{Dubovik:1975ch}: the toroidal dipole moment (TDM), $\tau(q^2)$.

The TDM and the anapole moment coincide in the case of $m_i = m_f$,
i.e. the incoming and outgoing particle are the same. This type of
static multipole moments does not produce any external fields in
vacuum but generate a free-field (gauge invariant) potential
\cite{Dubovik:1996gx}, which is responsible for topological effects
like the Aharonov-Bohm one.

\begin{figure}
  \centering
  \includegraphics[height=.25\textheight]{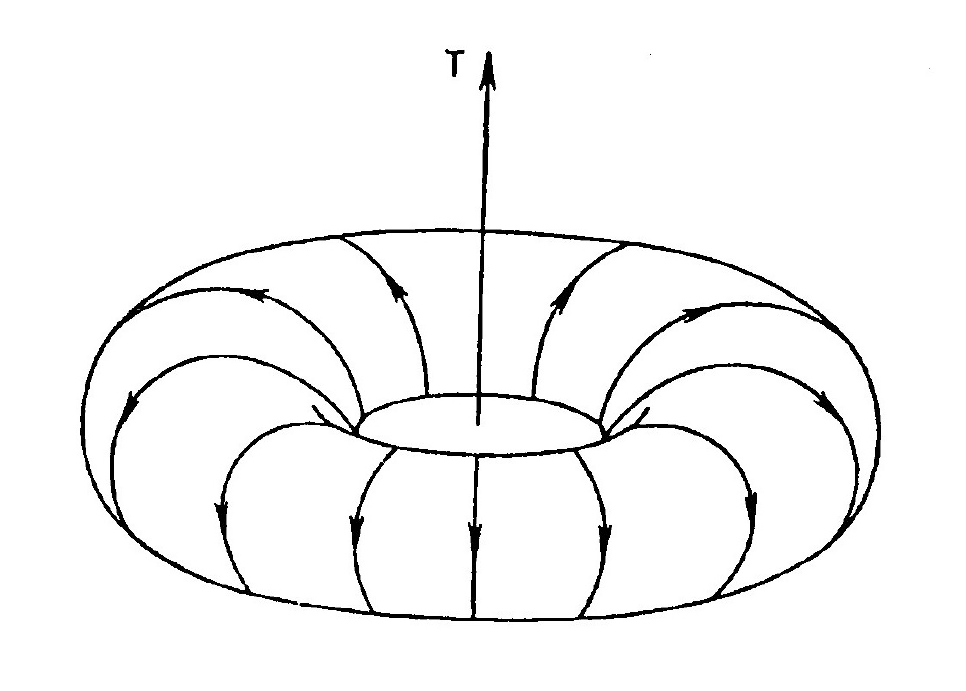}
  \caption{Current configuration with a toroidal dipole moment. The
    arrows on the torus indicate the direction of the current, and the
    TDM is directed towards the symmetry axis of the torus.}
\label{toro}
\end{figure}

The simplest TDM model (anapole) was given by Zel'dovich as a
conventional solenoid rolled up in a torus and with only one poloidal
current, see fig. \ref{toro}. For such stationary solenoid, without
azimuthal components for the current or the electric field, there
is only one magnetic azimuthal field different from zero inside
the torus.

\section{One-loop calculation}

\begin{figure}[h]
   \centering
   \includegraphics[scale=.25]{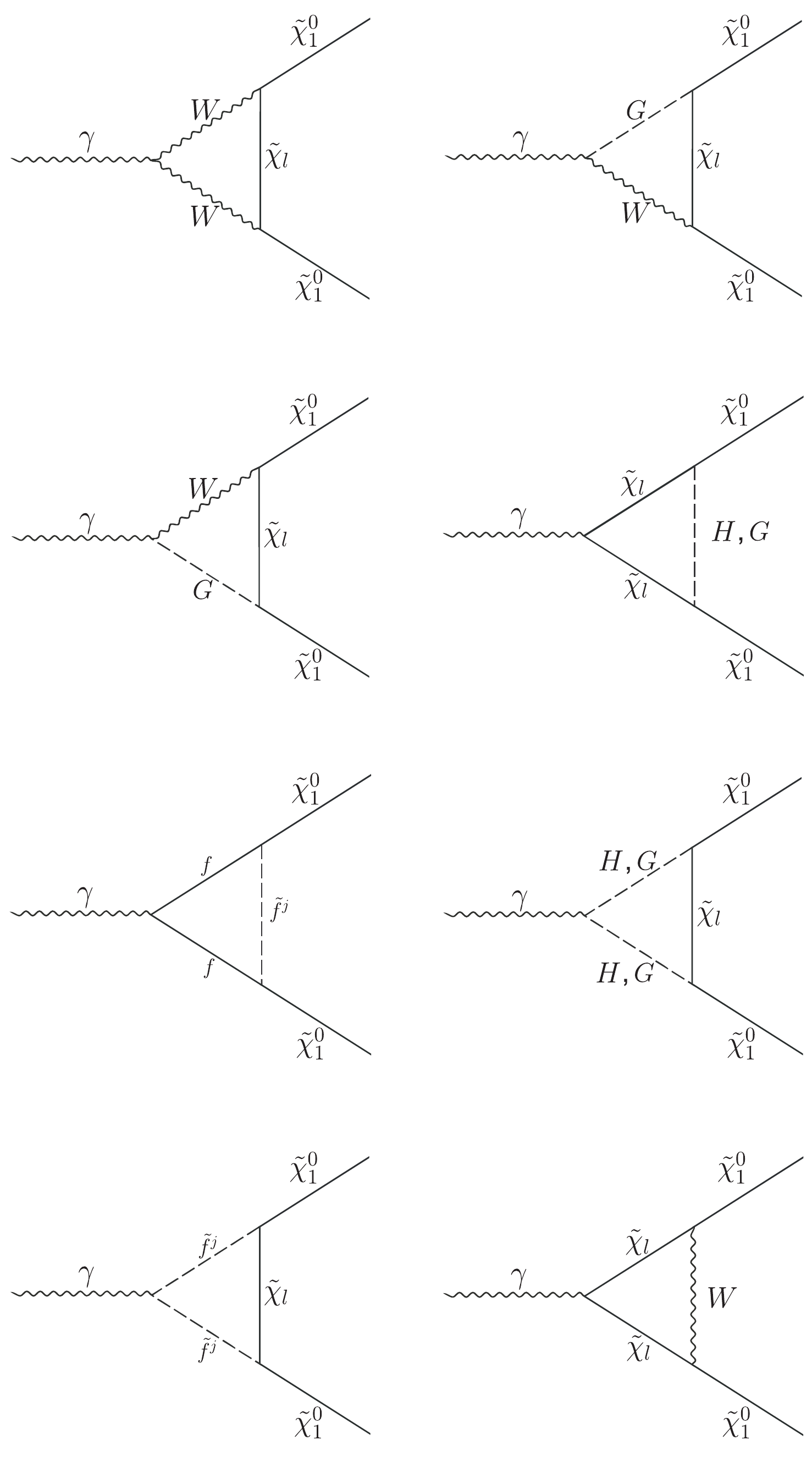}
   \caption{One-loop vertex corrections to the process
$\gamma\longrightarrow \chi_{1}^{0} \chi_{1}^{0}$.}
\label{one-loop-vertex}
\end{figure}

\begin{figure}[h]
   \centering
   \includegraphics[scale=.1]{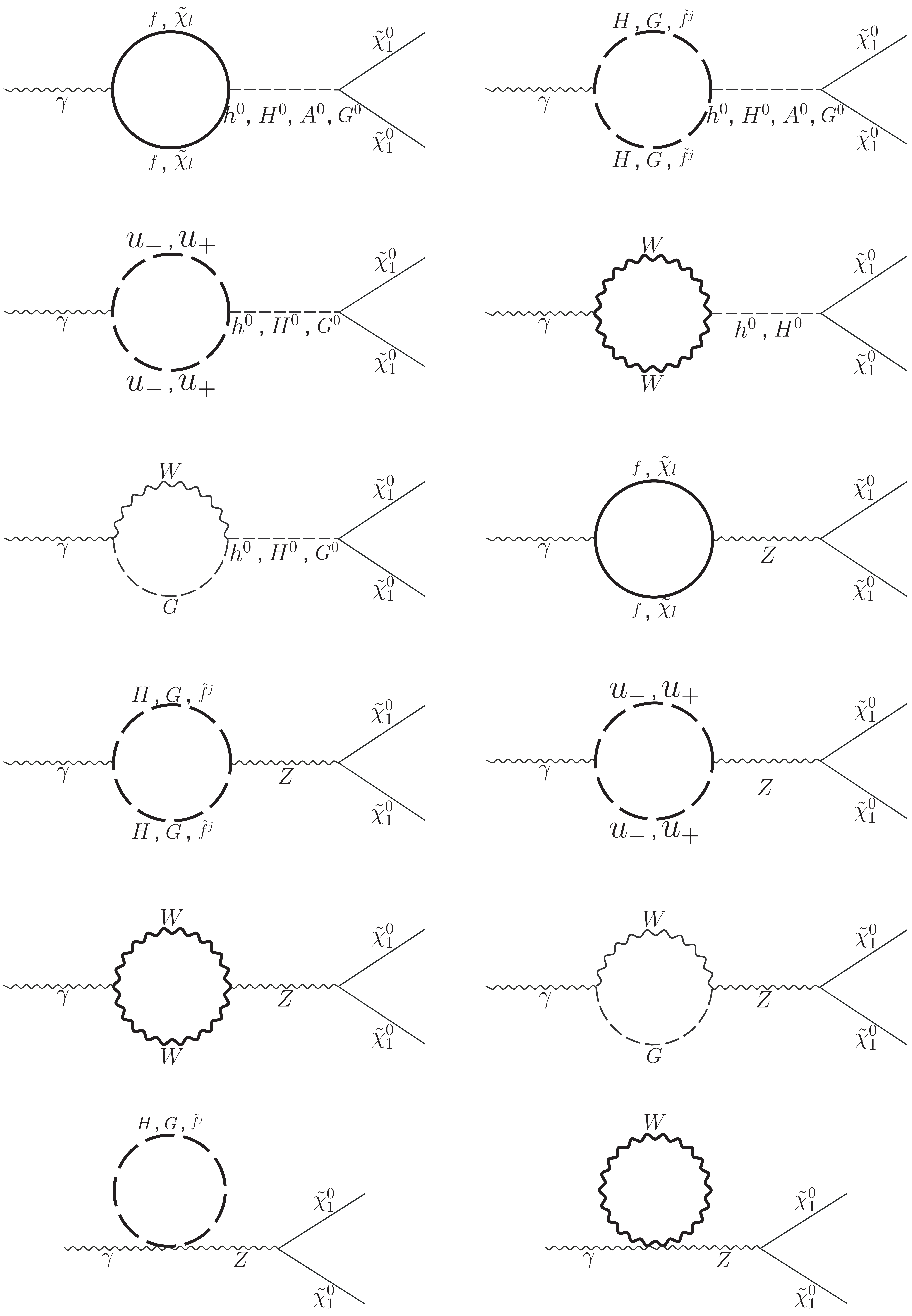}
   \caption{One-loop corrections to the self-energy for the process
     $\gamma\longrightarrow \chi_{1}^{0} \chi_{1}^{0}$.}
\label{one-loop-self}
\end{figure}

The TDM of the neutralino may be defined in the one-loop approximation
in the cMSSM by the Feynman diagrams shown in
figs. \ref{one-loop-vertex} and \ref{one-loop-self}, where $f$
represents the charged fermions of the SM.  Taking each fermionic
family separately we obtain 94 Feynman diagrams in total: 66
corresponding to self-energy and 28 to vertex corrections.

We use \textit{FeynCalc} to calculate the amplitude of these diagrams.
Since we are only interested in the terms that contribute to the
anapole form factor, we isolate the ones that have the Lorentz
structure $\gamma_{\mu}\gamma_5$. It is important to notice here that we work in the t'Hooft-Feynman gauge ($\xi = 1$). One of the first results we
obtain is that the self-energies $\gamma H^{0}$, $\gamma h^{0}$,
$\gamma A^{0}$ and $\gamma G^{0}$ do not contribute to the
calculation.  If we call  $\Xi_{i}$ the coefficient that multiplies 
$\gamma_{\mu}\gamma_5$ for the  \textit{ith} diagram, then we have
that

\begin{equation}
\sum_{i} \Xi_{i} = f_A (q^2 )q^2.
\end{equation}

To obtain the toroidal dipole moment $\tau=f_A(0)$ 
we use the l'Hopital rule and get

\begin{equation}
  \tau=f_A(0) = \lim _{q^{2}\rightarrow 0} \frac{\sum_{i} \Xi_{i}}{q^2} =
 \frac{\partial \sum_{i} \Xi_{i}}{\partial q^2} \mid _{q^{2}\rightarrow0}.
\end{equation}

\begin{figure}[h]
   \centering
   \includegraphics[scale=1]{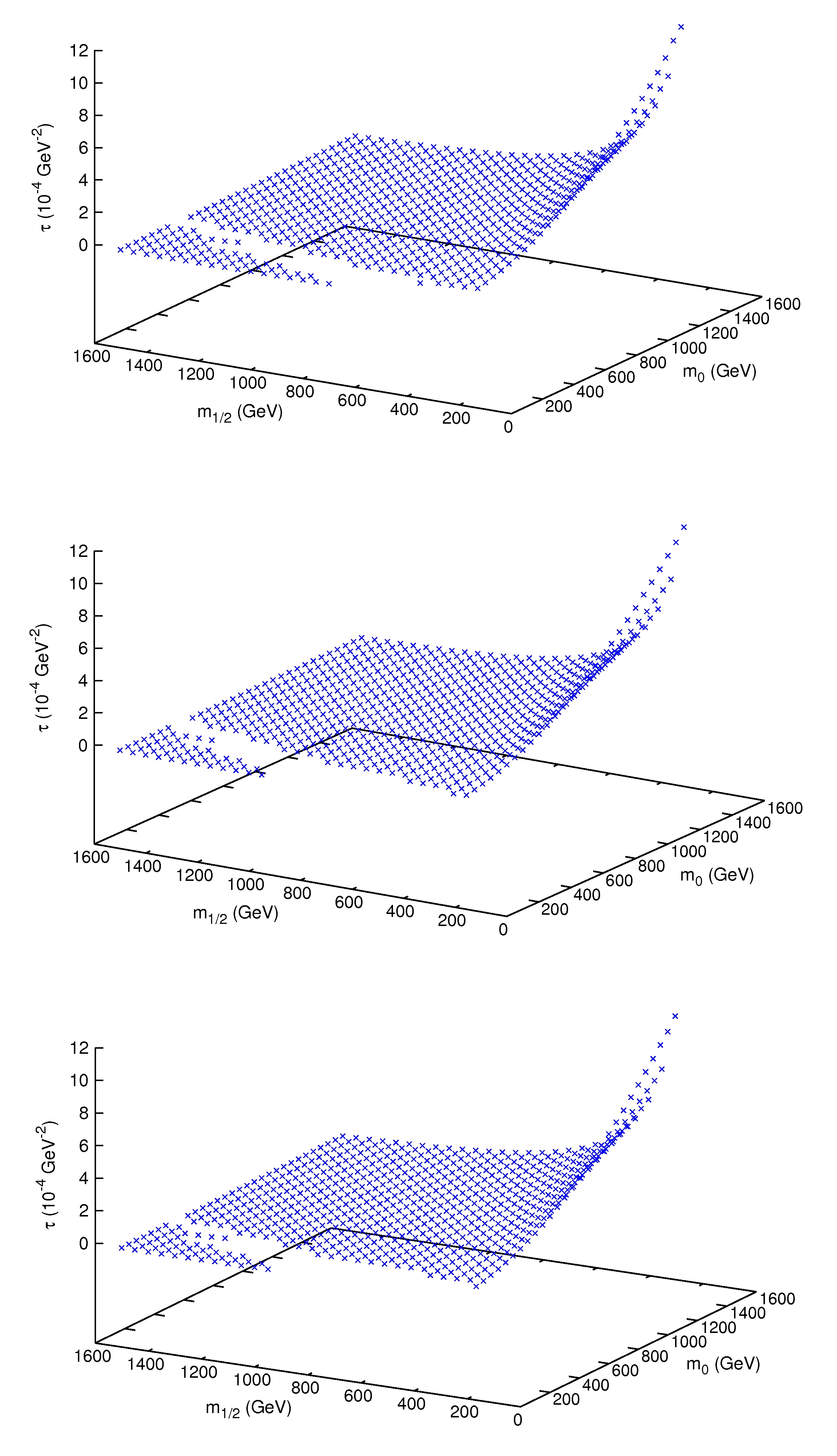}
   \caption{TDM for $\tan \beta=10$, ${\rm sign}\mu = +$ y $A_0 =
     -1000$ (top), $0$ (centre) y $1000$ (bottom) GeV.}
\label{10plus}
\end{figure}

\begin{figure}[h]
   \centering
   \includegraphics[scale=1]{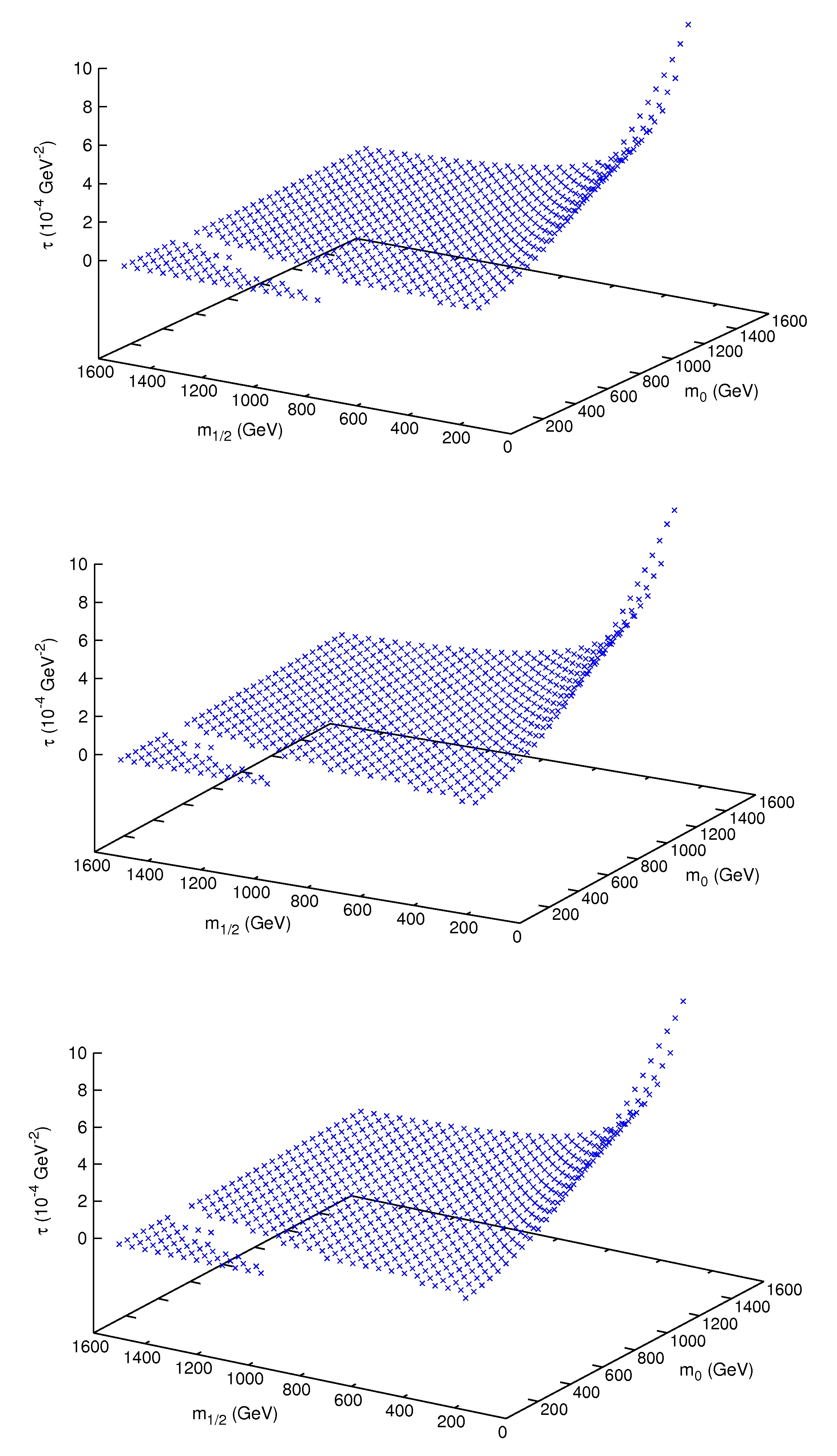}
   \caption{TDM for $\tan \beta=10$, ${\rm sign}\mu = -$ y $A_0 =
     -1000$ (top), $0$ (centre) y $1000$ (bottom) GeV.}
\label{10minus}
\end{figure}

\begin{figure}[h]
   \centering
   \includegraphics[scale=1]{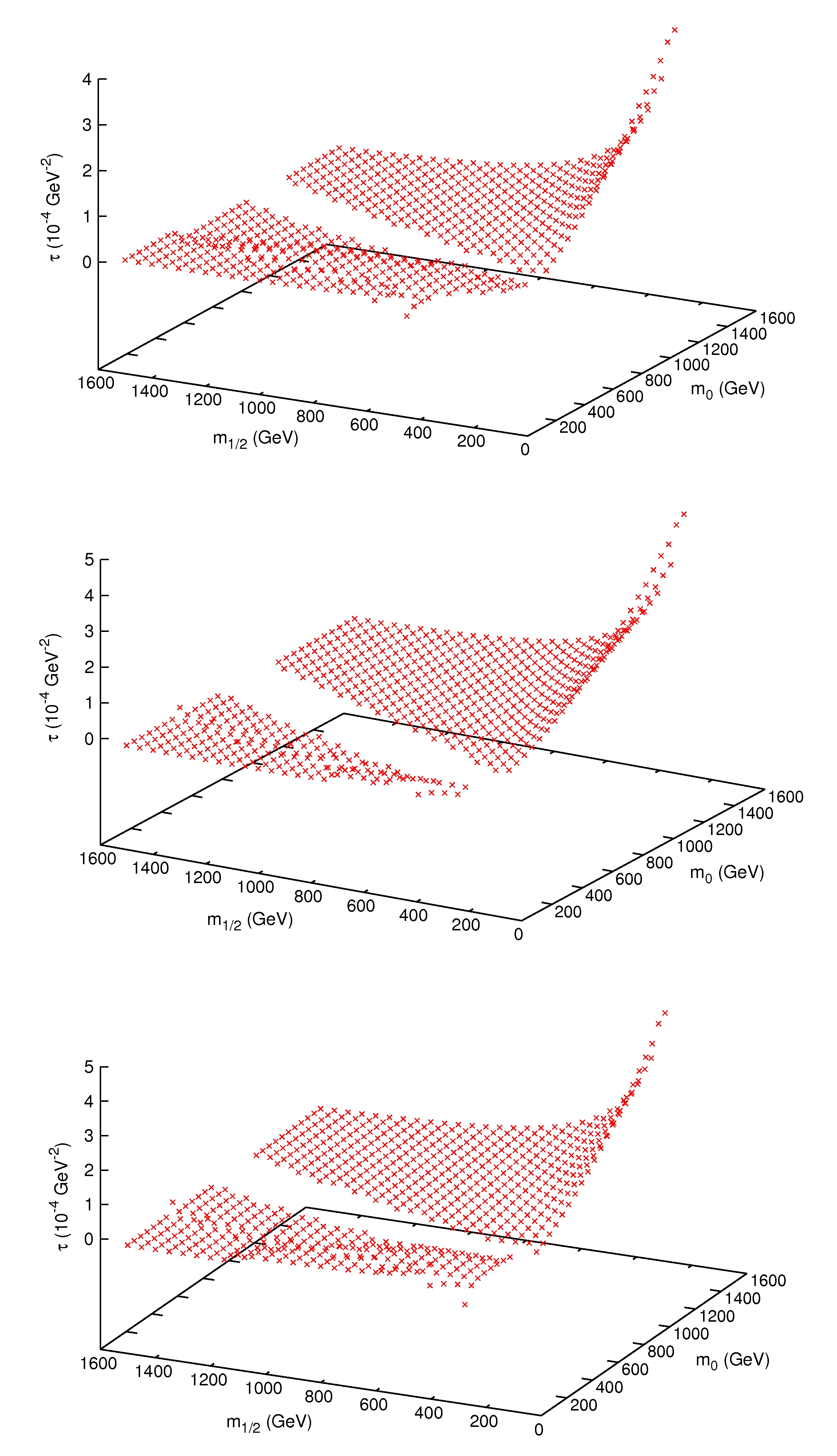}
   \caption{TDM for $\tan \beta=50$, ${\rm sign}\mu = +$ y $A_0 =
     -1000$ (top), $0$ (centre) y $1000$ (bottom) GeV.}
\label{50plus}
\end{figure}

The contributions to the self-energies have two point
Passarino-Veltman scalar functions of the type $B_{0} \left( q^{2},
  x^2, x^2 \right)$ and $B_{0} \left( 0, x^2, x^2 \right)$.  Likewise,
the contributions to the vertex corrections have two and three point
scalar functions of the type $B_{0} \left( q^{2}, x^{2},
  x^{2}\right)$, $B_{0} \left( M^{2}_{\tilde{\chi} _{1}^{0}}, y^{2},
  x^{2}\right)$ and $C_{0} \left( q^2, M^{2}_{\tilde{\chi} _{1}^{0}},
  M^{2}_{\tilde{\chi} _{1}^{0}}, x^{2}, x^{2}, y^{2} \right)$.  In
both cases $x$ and $y$ represent the masses of the particles in the loop.

When evaluating (6), derivatives of the Passarino-Veltman functions
appear.  To evaluate the $B_{0}$'s, as well as their derivatives, we
use \textit{LoopTools}\cite{Hahn:1998yk}. To evaluate the $C_{0}$'s and their
derivatives we expande them in a power series around $q^2 =0$ (see appendix).

The expression obtained for the toroidal dipole moment depends on
various parameters of the MSSM, including the supersymmetric particles
masses as well as the mass mixing matrix elements, the value of $\tan
\beta$, and the values of the soft breaking terms. We evaluate
the TDM within the cMSSM using \textit{Suspect}\cite{Djouadi:2002ze},
by fixing the value of $A_0$, $\tan \beta$ and ${\rm sign}\mu$, and
scanning over the other two parameters $m_0$ and $m_{1/2}$, from $0$ to
$1500$ GeV and $250$ to $1500$ GeV, respectively.

\begin{figure}[h]
   \centering
   \includegraphics[scale=.8]{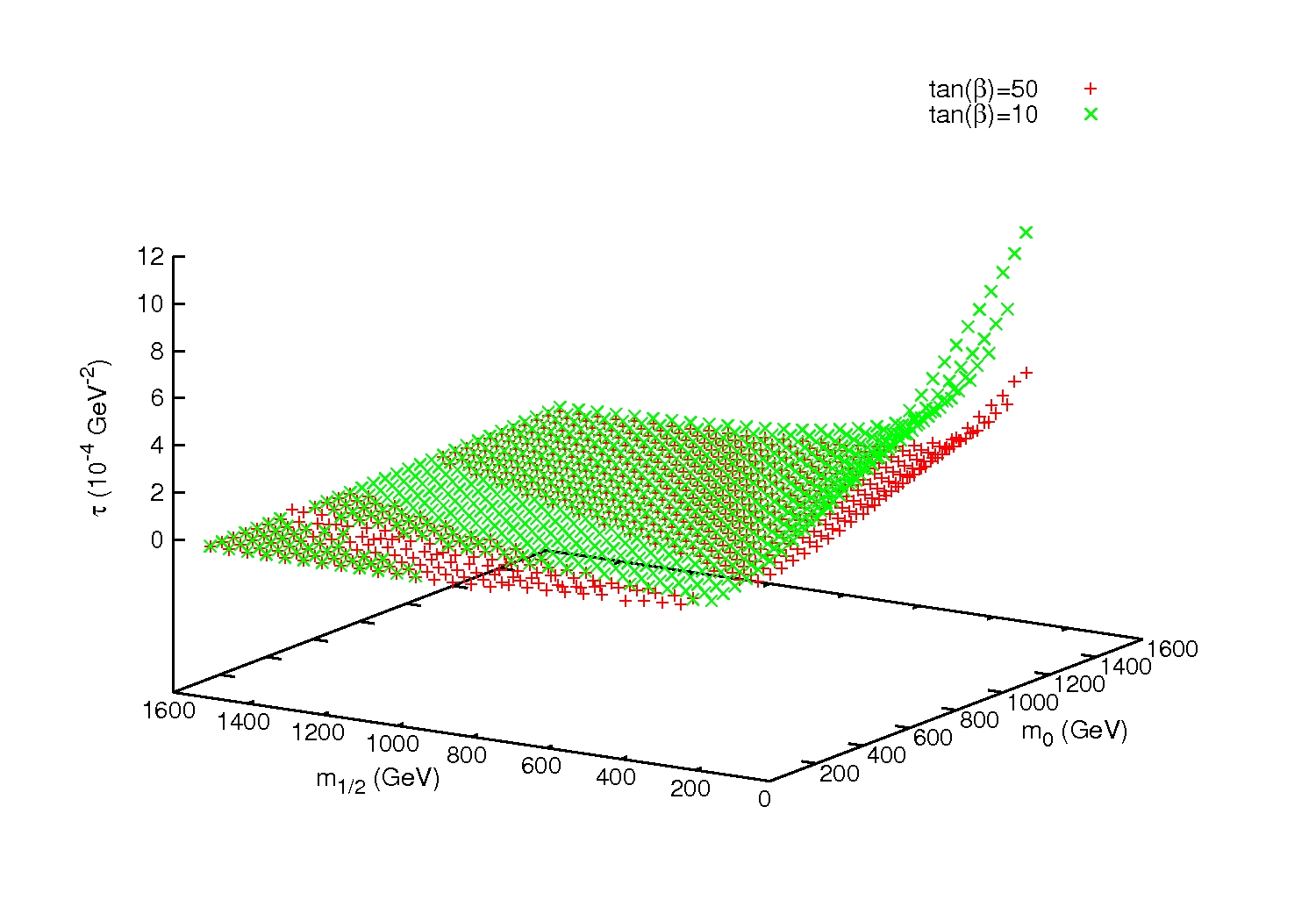}
   \caption{TDM for ${\rm sign}\mu = +$, $A_0 = 0$ GeV, $\tan \beta =
     10$ (green) and $\tan \beta = 50$ (red).}
\label{tanbeta}
\end{figure}

\begin{figure}[h]
   \centering
   \includegraphics[scale=.8]{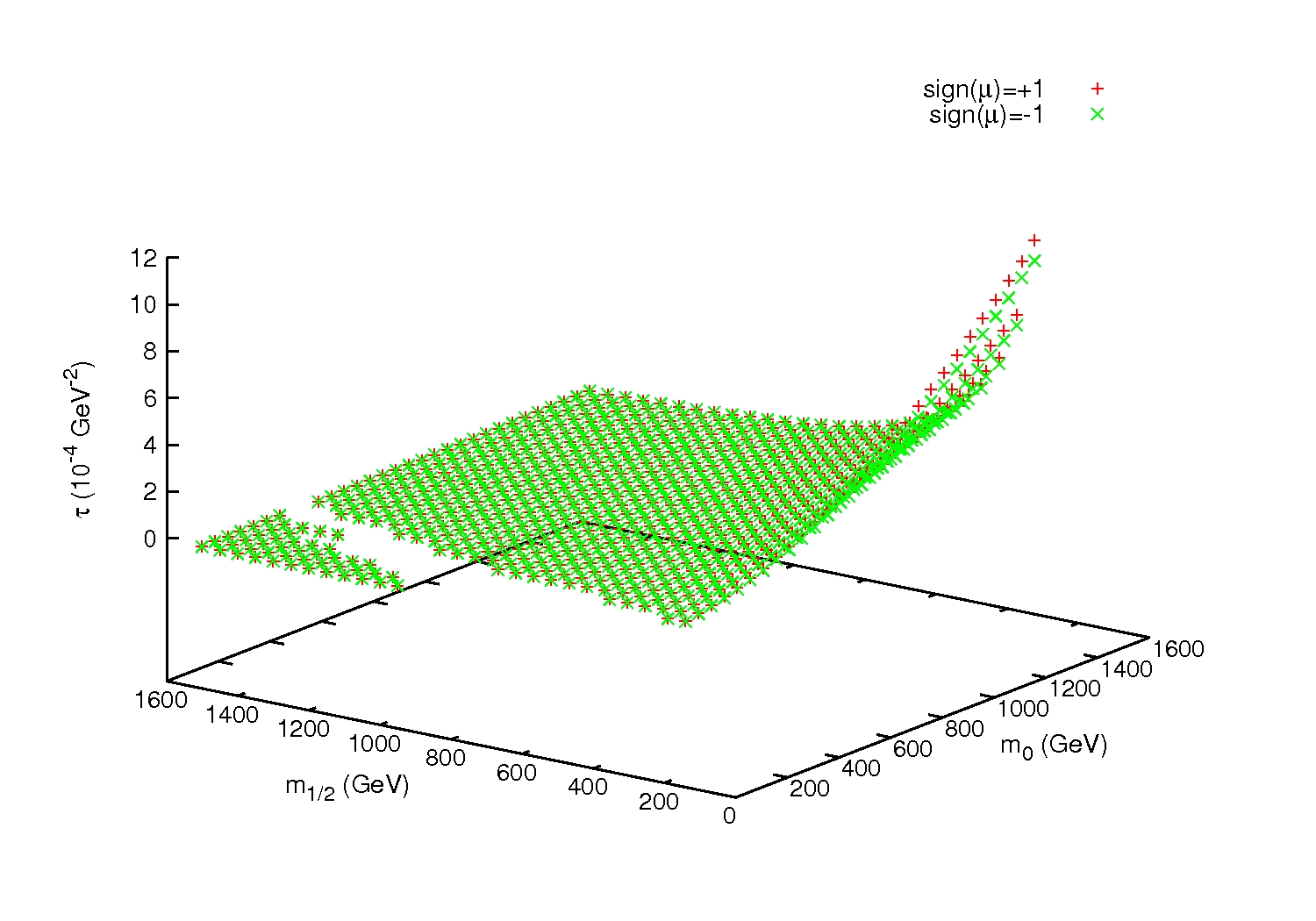}
   \caption{TDM for $\tan \beta = 10$, $A_0 = 0$ GeV, ${\rm sign}\mu =
     +$ (red) and ${\rm sign}\mu = -$ (green).}
\label{signmu}
\end{figure}

Figure \ref{10plus} shows the neutralino toroidal dipole moment for
$\tan \beta=10$, $\mu > 0$, and three different values of $A_0$,
$-1000$, $0$ and $1000$ GeV (top to bottom). Comparing the three
different plots, no dependence on $A_0$ is shown. The TDM is very low
for almost every region of the parameter space scanned, with values
between $10^{-5}$ and $10^{-8}$ GeV$^{-2}$. However the TDM increases
for increasing $m_0$ and decreasing $m_{1/2}$, reaching values over
$10^{-3}$ GeV$^{-2}$ for high $m_0$ ($\geq 800$ GeV) and low $m_{1/2}$
($\leq 400$ GeV). Similarly, figure \ref{50plus} shows the results for
$\tan \beta=50$, $\mu > 0$, and the same three different values of
$A_0$, $-1000$, $0$ and $1000$ GeV (top to bottom). The TDM reaches
values around $10^{-3}$ GeV$^{-2}$.

Figure \ref{10minus} shows the TDM for $\tan \beta=10$ and $\mu < 0$. Sign$\mu > 0$ may solve the problem of the discrepancy between the measured value of $g-2$ of the muon and the one predicted by the SM. However, this does not mean negative sign$\mu$ is ruled out since others mechanisms could solve this problem, and therefore sign$\mu<0$
should be taken into consideration.

Figure \ref{tanbeta} shows a comparison of two plots for different
$\tan \beta$ but same sign$\mu$ and $A_0$. This figure shows the
dependence of the TDM on $\tan \beta$. Figure \ref{signmu} shows a
comparison of two plots for different values of sign$\mu$ but same $\tan \beta$
and $A_0$. This figure shows no dependence of the TDM on sign$\mu$.

Notice that in all the plots the region for which $M_{\tilde{\chi}
  _{1}^{0}} = M_{\tilde{\tau}}$ is suppressed since we are not considering this possibility. This condition
($M_{\tilde{\chi} _{1}^{0}} = M_{\tilde{\tau}}$) separates the region
where the neutralino $\tilde{\chi} _{1}^{0}$ is the LSP and the one
where the stau $\tilde{\tau}$ is the LSP.

\begin{figure}[h]
   \centering
   \includegraphics[scale=.3]{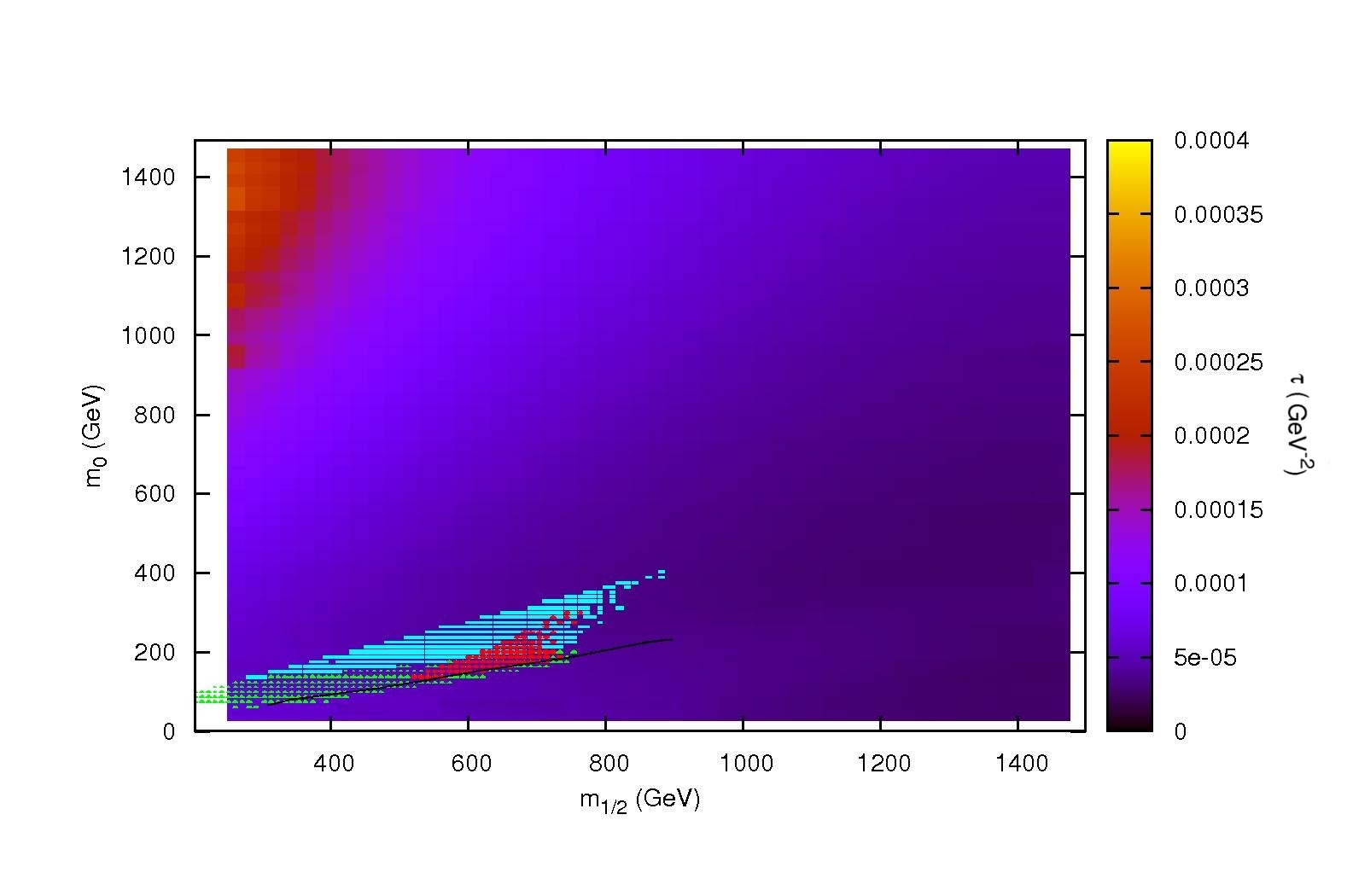}
   \caption{TDM results for $\tan \beta = 10$, $A_0 = 0$ GeV and ${\rm sign}\mu = +$ combined with restrictions from cosmological observational data (WMAP).}
\label{grafica}
\end{figure}

If combined with other criteria, our results can be used to further
reduce the parameter space of the cMSSM. As an example, figure
\ref{grafica} combines our results with those obtained in ref.
\cite{Stark:2005mp}. They studied the allowed parameter space for
diferent models (including cMSSM), supposing that the neutralino is
the main component of the CDM, varying $A_0$ and imposing the
restrictions of cosmological experimental data from WMAP. As can be
seen in the plot, their criteria favours a region with very low $m_0$,
which is compatible with a TDM $<10^{-5}$ GeV$^{-2}$ for the
neutralino. In other words, if a TDM is measured for a WIMP and it is
around or higher than $10^{-4}$ GeV, the neutralino of the cMSSM would
no longer be a good candidate for CDM in this region of parameter space.

\begin{figure}[h]
   \centering
   \includegraphics[scale=.30]{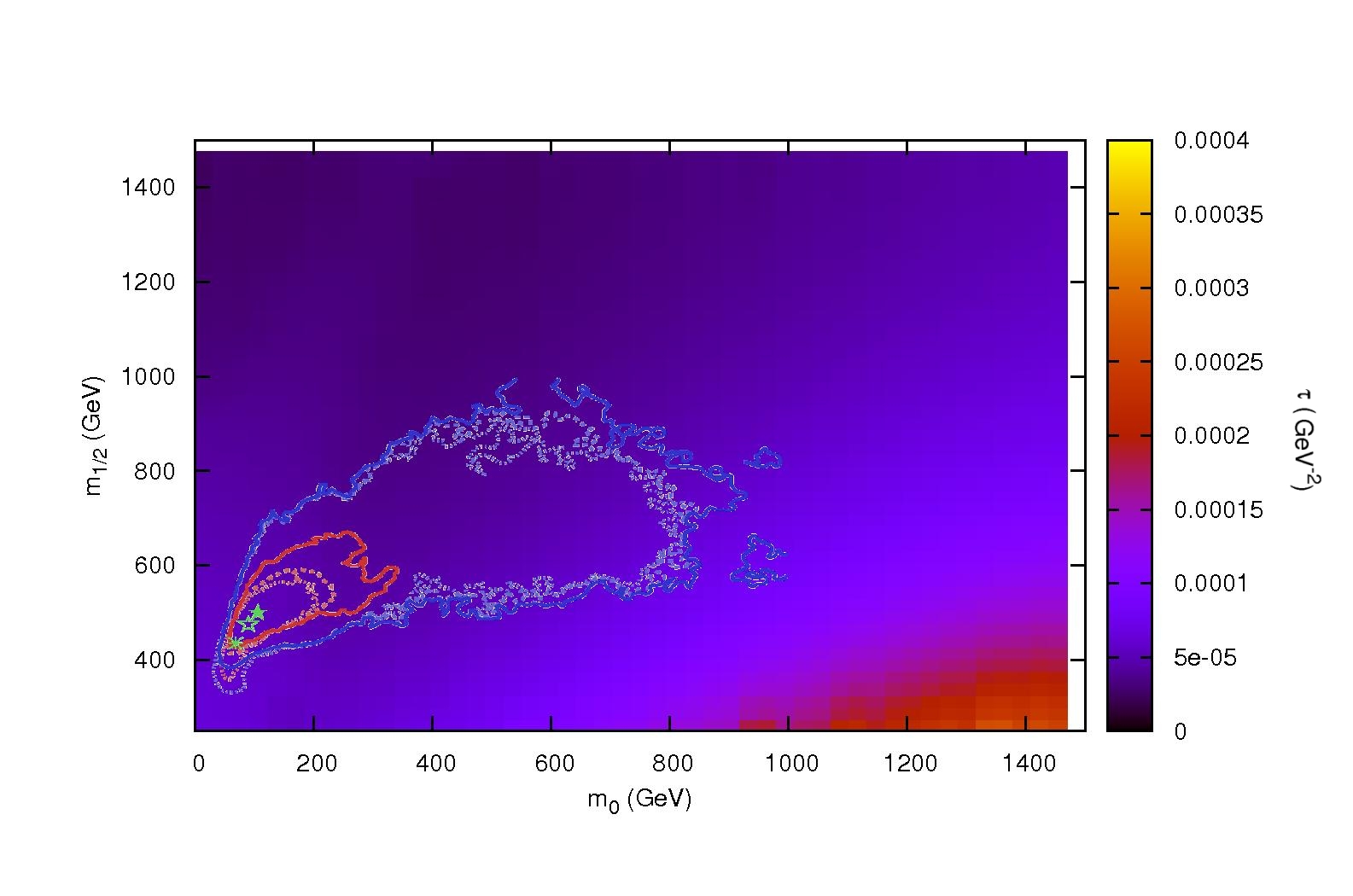}
   \caption{TDM results for $\tan \beta = 10$, $A_0 = 0$ GeV, ${\rm sign}\mu = +$ combined with restrictions from initial direct LHC searches for supersymmetry.}
\label{10plus0comp}
\end{figure}

In figure \ref{10plus0comp} we compare our results with a frequentist
analysis of the probable ranges of parameters of the MSSM using the
results of initial direct LHC searches for supersymmetry, combined with the required cold dark matter density, and the spin-independent dark matter scattering cross section 
\cite{Buchmueller:2011aa}. The figure shows the 68 and 95\% CL
contours (red and blue, respectively) both after applying the CMS and
ATLAS constrains (dashed and solid lines, respectively) and beforehand
(dotted lines), taken from their paper. Also shown in the figure as open (solid) green stars
are the best-fit points found after applying the CMS (ATLAS)
constrainsts in the cMSSM. As can be seen in the figure, this analysis
also favours a neutralino with a very low TDM.

\begin{figure}[h]
   \centering
   \includegraphics[scale=.95]{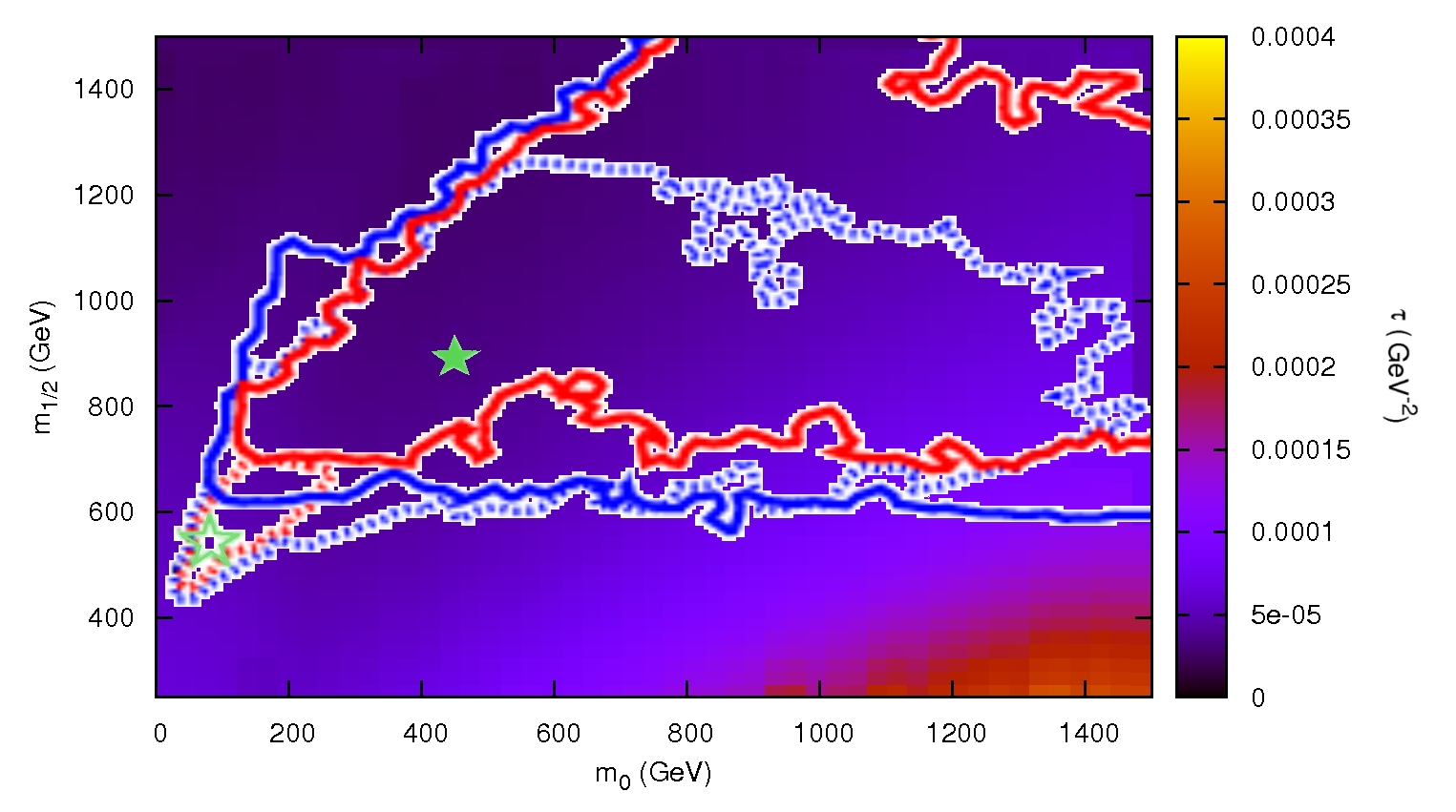}
   \caption{TDM results for $\tan \beta = 10$, $A_0 = 0$ GeV, ${\rm sign}\mu = +$ combined with restrictions from searches for supersymmetric signals using $~1/$fb of LHC data.}
\label{10plus04}
\end{figure}

In figure \ref{10plus04} we compare our results with another frequentist analysis of the CMSSM pararameter space including the public results of searches for supersymmetric signals using $~ 1/$fb of LHC data recorded by ATLAS and CMS and $~ 0.3/$fb of data recorded by LHCb in addition to electroweak precision and B-physics obsevables \cite{Buchmueller:2011sw}. The figure shows the 68 and 95\% CL contours (red and blue, respectively) with LHC$_{1/b}$ data (solid lines) and showing pre-LHC fits (dotted lines), taken from their paper. Although this analysis widens the allowed parameter space, it still favours a neutralino with very low TDM.

\section{Conclusions}

We calculated the only electromagnetic property of the lightest
neutralino: its toroidal dipole moment. Its characterization is
extremely valuable for discriminating different models which have the
neutralino as dark matter candidate. We performed the calculation in
the framework of the cMSSM, however a similar analysis can be performed for
other  models (work in progress). We found that the TDM of the neutralino is
highly sensitive to $m_0$, $m_{1/2}$ and $\tan \beta$, but very weakly
or practically non-dependent on $A_0$ and sign$\mu$.

All points in the parameter space we scanned give a TDM consistent
with the upper limit ($\sim10^{-2}$ GeV$^{-2}$) obtained by Pospelov
and ter Veldhuis \cite{Pospelov:2000bq} for WIMPs interacting with
heavy nuclei using data from the CDMS and DAMA experiments. However, this
data can and will be improved in the next few years helping to refine 
the upper limit, likely ruling out some regions of the parameter
space.

The TDM analysis can be used as another criteria to constrain the
parameter space of a given model which has a neutralino as candidate for dark
matter. Thus, according to our results, if a non-zero (around
$10^{-4}$-$10^{-3}$ GeV$^{-2}$) TDM could be measured for the
neutralino, that would indicate that the favored region of the
parameter space of the cMSSM would be high $m_0$ ($\geq 800$ GeV) and low $m_{1/2}$
($\leq 400$ GeV). Otherwise, other regions are compatible with a TDM
lower than $10^{-5}$ GeV.

If combined with other criteria (such as dark matter relic density, or
frequentist analysis of data from the LHC) the parameter space can be
reduced even further. In fact, the combination with these other criteria 
favours a region
with  low $m_0$, which is compatible with a $10^{-5}$-$10^{-8}$
GeV$^{-2}$ TDM for the neutralino. This means, among other things,
that if a TDM higher than $10^{-4}$ GeV$^{-2}$ is measured for a WIMP
it would exclude some regions of parameter space of the cMSSM (and other more
specific models), at least if the neutralino is the only component 
of dark matter. One should keep in mind that all analysis mentioned were made with this assumption.

\section{Acknowledgements}
We acknowledge very useful discussions with E. Ley Koo and A. Mondrag\'on.\\
This work was partially supported by UNAM grants PAPIIT IN111609 and IN113412.

\section*{Appendix: Scalar Three-point Function}

In this appendix, we analyse the Passarino-Veltman scalar three-point
function $C_0(q^2,x^2,x^2,z^2,z^2,y^2)$ \cite{Cabral:2000, Cabral:2006} which appears in the TDM
calculation. Here $q^2$ denotes the photon transfered 4-momentum, $x$
is the neutralino mass, and $y$ and $z$ are the masses of the
particles running in the loop.

\begin{figure}[h]
   \centering
   \includegraphics[scale=.3]{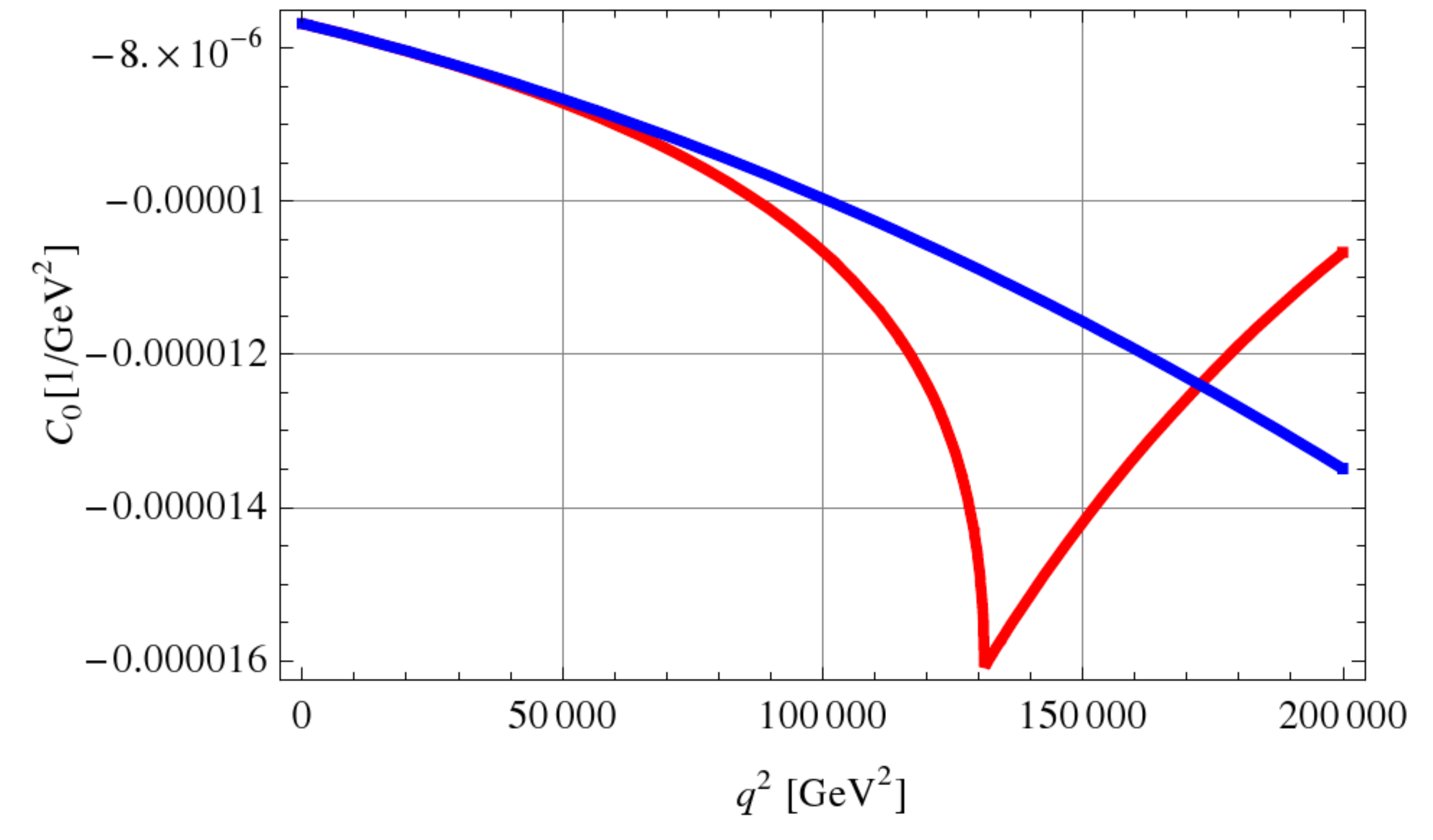}
   \caption{Comparison between numerical (red line) and approximate
     (blue line) scalar three-point function $C_0(q^2,x^2,x^2,z^2,z^2,y^2)$, with $x=97.7$ GeV, $y=415.4$ GeV and $z=80.43$ GeV. The analytical approximation (blue line) is only valid for $q^2\rightarrow 0$.}
\label{C0}
\end{figure}

The corresponding plot for this $C_0$ function can be seen in figure \ref{C0}.
The red line shows the numerical solution, the blue line
line represents the approximate solution, i.e., the Taylor expansion
around $q^2 = 0$, which can be written as follows:

\begin{equation}
C_0\left( q^{2},x^{2},x^{2},z^{2},z^{2},y^{2}\right) = \alpha_{0} +
\alpha_{1} q^{2} + {\cal O}(q^4).
\end{equation}

\noindent 
The coefficients $\alpha_i$ are functions of the masses:

\begin{equation}
\alpha_0 = \frac{\log\left( \frac{y^2}{z^2}\right)}{2 x^2} + a\log\omega, 
\end{equation}

\begin{equation}
\alpha_1 =
\frac{x^4-y^2x^2-2z^2x^2+z^4-y^2z^2}{6x^2z^2(-x+y-z)(x+y-z)(-x+y+z)(x+y+z)} 
+ \frac{\log\left( \frac{y^2}{z^2}\right)}{12 x^4}+ b\log \omega,
\end{equation}

\noindent where

{\scriptsize
\begin{equation}
\omega = \frac{\left(
  ix^2+iy^2-iz^2+\sqrt{-y^4+2(x^2+z^2)y^2-(z^2-x^2)}\right)
\left( ix^2-iy^2+iz^2+\sqrt{-y^4+2(x^2+z^2)y^2-(z^2-x^2)}\right) }
{\left( -ix^2+iy^2-iz^2+\sqrt{-y^4+2(x^2+z^2)y^2-(z^2-x^2)}\right)
\left( -ix^2-iy^2+iz^2+\sqrt{-y^4+2(x^2+z^2)y^2-(z^2-x^2)}\right)},
\end{equation}
}

\begin{equation}
a= \frac{i(x^2+y^2-z^2)}{2x^2\sqrt{-x^4+2y^2x^2+2z^2x^2-y^4-z^4+2y^2z^2}}
\end{equation}

\noindent and

{\scriptsize
\begin{equation}
b= \frac{i(x^2+y^2-z^2)(x^4-4y^2x^2-2z^2x^2+y^4+ź^4-2y^2z^2)}
{12x^4(-x+y-z)(x+y-z)(-x+y+z)(x+y+z) \sqrt{-x^4+2y^2x^2+2z^2x^2-y^4-z^4+2y^2z^2}}.
\end{equation}
}

\bibliography{biblio}{}
\bibliographystyle{h-physrev4}


\end{document}